# A Modern Analysis of Aging Machine Learning Based IoT Cybersecurity Methods

Sam Strecker[1], Rushit Dave[1,*], Nyle Siddiqui[1], Naeem Seliya[1]

[1]Department of Computer Science, University of Wisconsin – Eau Claire, Eau Claire, US
*Corresponding author: daver@uwec.edu

**Abstract** Modern scientific advancements often contribute to the introduction and refinement of never-before-seen technologies. This can be quite the task for humans to maintain and monitor and as a result, our society has become reliant on machine learning to assist in this task. With new technology comes new methods and thus new ways to circumvent existing cyber security measures. This study examines the effectiveness of three distinct Internet of Things cyber security algorithms currently used in industry today for malware and intrusion detection: Random Forest (RF), Support-Vector Machine (SVM), and K-Nearest Neighbor (KNN). Each algorithm was trained and tested on the Aposemat IoT-23 dataset which was published in January 2020 with the earliest of captures from 2018 and latest from 2019. The RF, SVM, and KNN reached peak accuracies of 92.96%, 86.23%, and 91.48%, respectively, in intrusion detection and 92.27%, 83.52%, and 89.80% in malware detection. It was found all three algorithms are capable of being effectively utilized for the current landscape of IoT cyber security in 2021.

*Keywords:* Internet of Things, Machine Learning, Cybersecurity, Intrusion Detection, Malware Detection, Botnet Attacks

## 1. Introduction

The number of Internet of Things (IoT) devices connected to the Internet has exceeded fifty billion devices since 2020 [24]. This results in a dire need to continually advance the cyber security field as stated in [1]. Previously vetted and approved security techniques need to be continually re-tested with new datasets. New and emerging technologies may deviate from past implementations and introduce new outlets for malicious users to exploit. For example, Google's Home Mini uses a hard-coded DNS to try and deter users from DNS based adblocking, but the result is a new security threat. The device overrides the DNS set on the local network, and in this case uses Google's DNS (8.8.8.8/8.8.4.4). Provided more information, a malicious user can now exploit the device with a current zero day named *'Name:Wreck'*. This is simply one example of thousands that cyber security analysts need to be aware of first before it is in the hands of a malicious user. This article will compare the effectiveness of three different machine learning based cyber security techniques on the Aposemat Iot-23 dataset published in January of 2020 [11]. Further, we produce an analysis of aging machine learning based IoT cybersecurity methods on newly published datasets that bring forth never-before seen IoT devices and their accompanying technologies/protocols, as seen in [5]. We test RF, SVM, and K-Nearest Neighbor algorithms to be used for IoT cyber security in the present year of 2021. Section 2 will discuss previous application of IoT and machine learning in academic literature, Section 3 will discuss in detail the dataset and algorithms we use in this article, Sections 4 and 5 serve as an exhibition and analysis of our observed results, and Section 6 summarizes and concludes our work.

## 2. Background

The phrase "Internet of Things" has been around for over two decades. Created in 1999, [16] initially used the term to refer to the use of radio-frequency identification tags in an assembly line. Today the phrase IoT is more encompassing and includes any device that has the ability to collect data and communicate this data via the internet [3, 6, 8]. Due to its usefulness and versatility, IoT devices became deeply integrated into the workings of society. From healthcare, to e-commerce, to one's own living room, IoT devices are abundant in presence. For instance, the city of Padua, in Italy, utilizes IoT networks to monitor carbon monoxide levels, traffic flow, noise levels, streetlights, power usage, and much more as stated in [19]. With the rise of technological prowess, so does the sophistication of malicious attacks on these technologies, thus making the need for stronger cyber security methods imperative [9]. New methods to protect IoT devices from malicious users are continuously being developed, however, developing well-made cyber security methods for IoT devices comes with its own unique challenges. IoT devices are very resource limited, similar in device signature, prone to botnet attacks, and can be found in edge computing scenarios. This results in the necessity for unique cyber security that can accommodate these limitations. Researchers in [8] and [20] studied the effectiveness of (CoAP) - the constrained application protocol - which presented a cost-effective solution to protect data transmission within IoT networks in real-time. Through the optimization of this protocol, what would have normally been unachievable with current cyber security techniques was proven possible in [8] and [20] with accuracy rates reaching a peak of 97%.. Overall, IoT cybersecurity could be greatly improved when compared to other areas like anti-fraud for online transactions [21]. With the ever-increasing number of IoT devices on the internet, the occurrences of botnet attacks are growing proportionally to the volumetric traffic of each attack in accordance with [9]. [10] exhibited the power of botnets when a peak Distributed Denial of Service (DDOS) attack with a bandwidth of 1.1Tb/s targeted the cloud service company Cloudflare, pushing it over the limit to what Cloudflare could mitigate resulting in large network outages. Current IoT security techniques rely on three main machine learning algorithms for device identification, intrusion attacks, and malware identification: SVMs, RFs, and KNNs. In essence, these three algorithms can be effectively utilized as watch dogs on IoT networks to prevent attacks in real-time. This study examines the effectiveness of RFs, SVMs, and KNNs for IoT cybersecurity against modern datasets.

According to [1], the use of supervised learning techniques like SVM, KNN, RF, Naïve Bayes, and artificial neural networks (ANN) are effective in identifying network traffic of IoT devices. More specifically, they have the capability to detect network intrusions as well as spoofing attacks. To identify Denial of Service (DoS) attacks, the use of multivariate correlation is needed. It extracts the geometrical correlations of network traffic features making the model 92% more accurate. Excluding deep learning, the researchers concluded that the RF performed most optimally for malware detection and KNN performed best for network intrusion. In [4], researchers set out to find a machine learning algorithm that had the ability to detect DDOS attacks, also seen in [7]. They chose to study KNN, SVM with Linear Kernel, Decision Tree (DT), and RF. The dataset was created from three IoT devices from ten minutes of captured network traffic. The models were trained using 85/15 split training and used the Sci-kit learn Python library. The researchers concluded stateless features were more beneficial to classification than stateful features as well as observing RF performing the best and the SVM the worst [4]. Similarly in [15], the researchers found the SVM performed the worst in comparison to KNN and RF [12]. In the context of binary classification, they were able to achieve accuracies over 99% with a RF. The researchers also found the Radial Basis Function kernel for SVMs worked almost twice as well as the Linear kernel. Following the same trend in [13], researchers utilized Google's MapReduce as a backbone for network traffic feature extraction, translation, and analysis of changing network features. They tested seven different machine learning based algorithms and RF performed the best with a precision of 0.9994 with SVM trailing far behind at 0.7714. Researchers in [13] set out to test the effectiveness of five different algorithms for intrusion detection. They sourced a dataset from Kaggle and selected 8 different feature vectors to learn from. Training was complete with 80/20 split testing as well as five-fold cross validation referenced in [14]. They found RF performs optimally with larger datasets opposed to SVM which regressed in performance. The researchers concluded RFs, Decision Trees, and KNNs could all sufficiently classify and differentiate between normal and attack data and found average accuracy rates observed in the field of IoT to be around 90% for most machine learning algorithms.

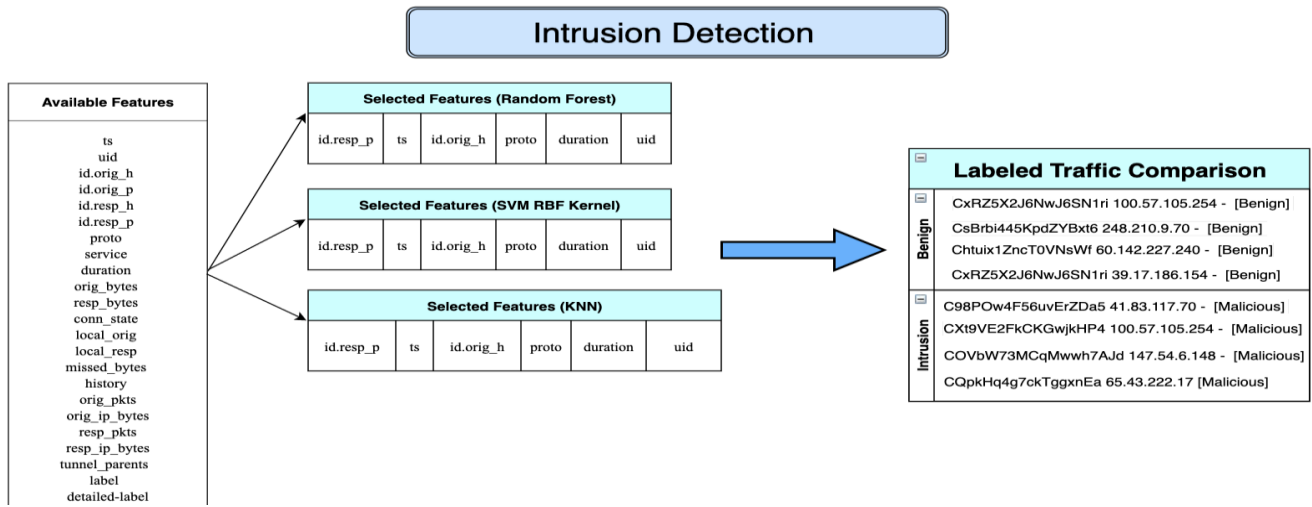

**Figure 1.** Feature selection for intrusion detection

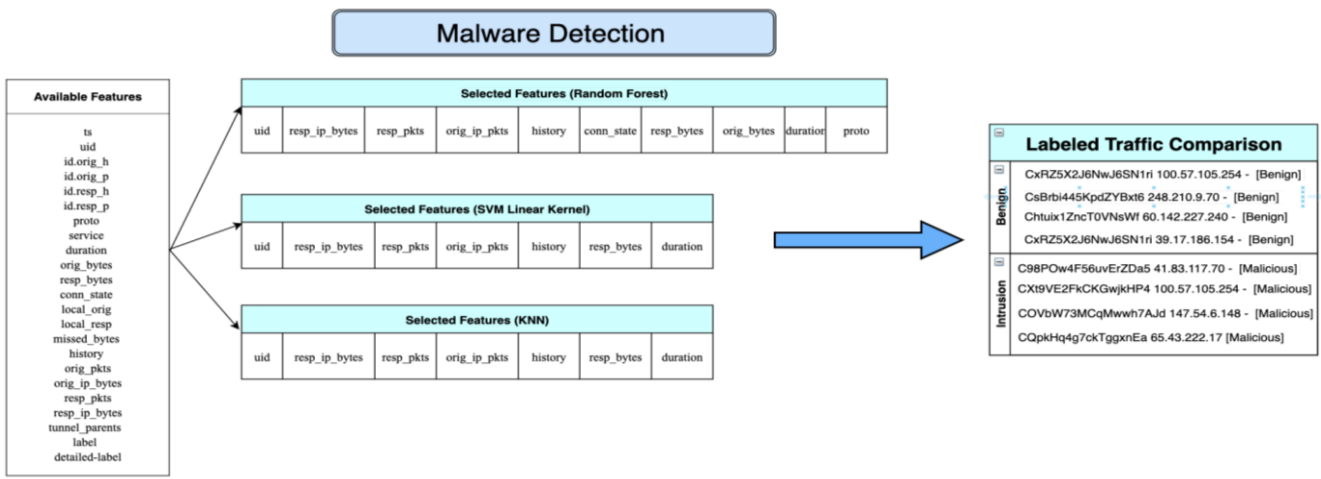

**Figure 2.** Feature selection for malware detection

## 3. Methodology

The first step in searching for a dataset was verifying it met the criteria of being published within a year of 2021. After comparing many datasets, our study concluded the *'Aposemat Iot-23'* dataset sufficiently met all our criteria. Modern datasets are integrable to the quality of results as recently published datasets will include newer devices which current cyber security systems haven't interacted with before. This leads to untested circumstances and uncovers new exploit surfaces. The study injected the dataset to a Pandas (v1.2.4) framework within the Spyder IDE (v4.2.3) running the Scikit-learn toolkit (v0.24). Training was performed on a PC with the following specifications: Windows 10 (10.0.18363), Intel i7-6800K, 32Gb RAM, and RTX2070 (driver version DCH 466.11). The SVM was tested with both RBF and Linear kernels to account for any variations or performance increases. The study implemented holdout validation with 60/40, 70/30, and 80/20 splits for training the data for both malware detection and intrusion detection to verify the validity of our trained models. The provided Receiver Operating Characteristic (ROC) curves and confusion matrixes are the averaged results from each split.

This article optimized the feature sets for each algorithm for both malware and intrusion detection to get the best performance from each algorithm. To optimize the data, the process started by assessing all the available features in our labeled traffic captures as pictured in Figure 1 and 2. Some features like missed_bytes or resp_pkts were not useful in identifying malware or intrusion-based attacks so they were immediately dismissed for testing. This is due to the fact that these features do not contribute to the probability of identifying a malware or intrusion-based attack with simpler machine learning algorithms. However, note that inclusion of these features may become useful with deep learning algorithms or more advanced datasets that can leverage more data points. Each algorithm shared the same base set of capture features and some algorithms like the RF worked better with an additional feature added. In the case of malware detection, the RF had lower false positive rates and higher accuracies when the features proto and orig_bytes were added (Fig. 2). However, the *proto* feature increased false positive rates significantly in the KNN and SVM as malware and non-malicious programs often use the same UDP and TCP protocols. Additional testing should train with Peer2Peer based traffic and anonymized MAC addresses which has become common practice within the field of cybersecurity.

### 3.1 Malware Detection

To properly identify malware on the local IoT network, our testing utilized a base feature set of uid, resp_ip_bytes, resp_pkts, orig_ip_pkts, history, resp_bytes, and duration as noted in Figure 1. The uid feature allowed the researchers to track a device across the local network regardless of if the device's IP changes or techniques to spoof or mask the device's identity occur. The resp_ip_bytes and resp_bytes sizes are typically very small or zero when malware is present on the network. Contrary to resp_ip_bytes, certain malware increases the orig_ip_pkts total, making this a good identifying metric. As for the history feature, it had minimal to slight effects on lowering false positive rates. Since the results were not negatively impacted, this article kept it as a feature as future testing may utilize it. Finally, the duration is a great feature to use as malware generally has very long connection times. A benign connection may have a connection length of 0.001482 seconds while malware is at 3.151458 seconds, giving a clear flag of suspicious activity.

### 3.2 Intrusion Detection

Similarly to malware detection, intrusion detection uses the uid to better track the device across the local IoT network. They also share the duration feature to identify suspicious lengths of time indicative of malicious intent. The timestamps (TS) feature is unique to intrusion-based attacks as a vertical or horizontal port scan will have very similar TS from the same device across many IPs or ports on the local network, depending on attack type. To better utilize the TS feature, it needs to be paired with an identifier for cross referencing. In this case, the id.resp_p and if.orig_h features can be used to determine the IP address of the origin device and IP of the destination device. This contributed to higher accuracies as similar timestamps can be compared against host and receiving IP addresses with connection duration to produce significant results.

Table 1. Intrusion detection averages

| Algorithm | Accuracy | F1 | Recall | TP | FN |
|---|---|---|---|---|---|
| RF | .9296 | .9588 | .9403 | .9503 | .0318 |
| KNN | .9148 | .9416 | .9293 | .9191 | .0356 |
| SVM | .8623 | .8671 | .8614 | .8623 | .0726 |

## 4. Results and Analysis

This study found the RF algorithm to be the best performing as seen in Table 1. It boasted high accuracies of 92.96% and 92.27% in intrusion and malware detection, respectively, and lowest overall false positive rates. The SVM was found to perform more optimally for malware detection with the Linear kernel and for intrusion detection with the RBF kernel. The KNN was the second-best performing algorithm, following closely behind RF in intrusion detection and malware detection with F1-scores within 3% of the RF and accuracies of 91.48% and 89.80%. With the exclusive features selected in this article, we were able to identify horizontal and vertical port scans as well as malware on an IoT network. The authors of the dataset labeled the traffic as benign or malicious, so this article was able to validate our results for accuracies and false negatives rather than relying on timely lookups that are prone to human error. Our research also concluded that the RF had a more consistent ROC curve across both scenarios compared to the KNN and SVM. The underlying factors to contribute to this are not directly known as the SVM showed irregular inconsistencies, potentially from the different kernels used between malware and intrusion detection as highlighted in [14]. Through various testing and validation methods, this study was able to conclude that aging machine learning algorithms are still effective at identifying malware and intrusion detection for IoT devices and networks. Tables 1 through 2 exhibit the averaged results for these algorithms.

Table 2. Malware detection averages

| Algorithm | Accuracy | F1 | Recall | TP | FN |
|---|---|---|---|---|---|
| KNN | .8980 | .9280 | .8971 | .8982 | .0411 |
| RF | .9227 | .9393 | .9330 | .9193 | .0372 |
| SVM | .8352 | .8559 | .8298 | .8352 | .0502 |

Table 3. Malware detection comparison

| Algorithm | Methodologies | Results | Pros | Cons |
|---|---|---|---|---|
| RF | Identify horizontal port scans by observing port, time, IP/UID, protocol, and service | RF was the best performing algorithm | Lowest false negative rates | 60:40 training F1 score was 0.9313 |
| KNN | Identify horizontal port scans by observing port, time, IP/UID, protocol, and service | KNN was the second-best performing algorithm | High recall rates | Too many features can lower accuracy |
| SVM (RBF) | Identify horizontal port scans by observing port, time, IP/UID, protocol, and service | SVM had an average recall of 0.856 | High precision and F1 scores | Recall could be improved upon |

Table 4. Intrusion detection comparison

| Algorithm | Methodology | Results | Pros | Cons |
|---|---|---|---|---|
| RF | Detect malware with 9 distinct network features | Average F1 score of 0.9415 | Best performing with highest accuracies | 60:40 False negative could be improved |
| KNN | Feature set reduced by 3 to improve recall and accuracy | High average ROC curve area of 0.881 | High F1 and low false negatives | Feature set needed to be reduced to improve accuracy |
| SVM (Linear) | Detect malware with feature set of size 6 to improve accuracies and recall | Average accuracy of 88.61% | High average true positive of 86.26% | Highest false positive rate |

Tables 3 and 4 highlight a low-level comparison of the three algorithms tested, all of which have the combined averages of the 60/40, 70/30, and 80/20 data splits.

## 5. Discussion

As [1, 2] previously found, the RF was the best performing algorithm out of the three tested. The confusion matrices and ROC curves of our results are seen in Figures 3 and 4. The visualization of our results assists in better understandings of the data and its granularity. The ROC curve measures the ratio between the true and false positive rates, with higher ROC values indicating the algorithm can effectively distinguish between positive instances in the data; in the context of our experiment, if a device is benign or malicious. A comparison between the area below the ROC curve for the SVM, KNN and RF can also be seen in Table 6. As stated before, the numbers present are averages of 60/40, 70/30, and 80/20 data splits to reduce bias.

These results in Table 2, 3, and 4 support the observations from [12, 17] that the differences in kernel had a large impact on the results and that SVMs perform the worst in intrusion detection. We observed similar results in our experiment as [4], where the authors successfully created a system that can identify malicious or anomalous data in real-time and demonstrated how RFs outperformed the other for algorithms tested. This is also seen in [13] and [18]. Similarly to our results, [4] also demonstrated the RFs optimal performance when fed stateless feature sets only. [13] and [18] were also able to accurately identify attack types based upon the data and identify horizontal and vertical port scans with the feature sets, which we successfully replicated. This is observed in the high true positive rate and low false negative scores across all of our algorithms in our confusion matrices. The determination can be safely made that each algorithm is suitable for modern-day malware and intrusion detection, an improved and refined conclusion to results in [22, 23].

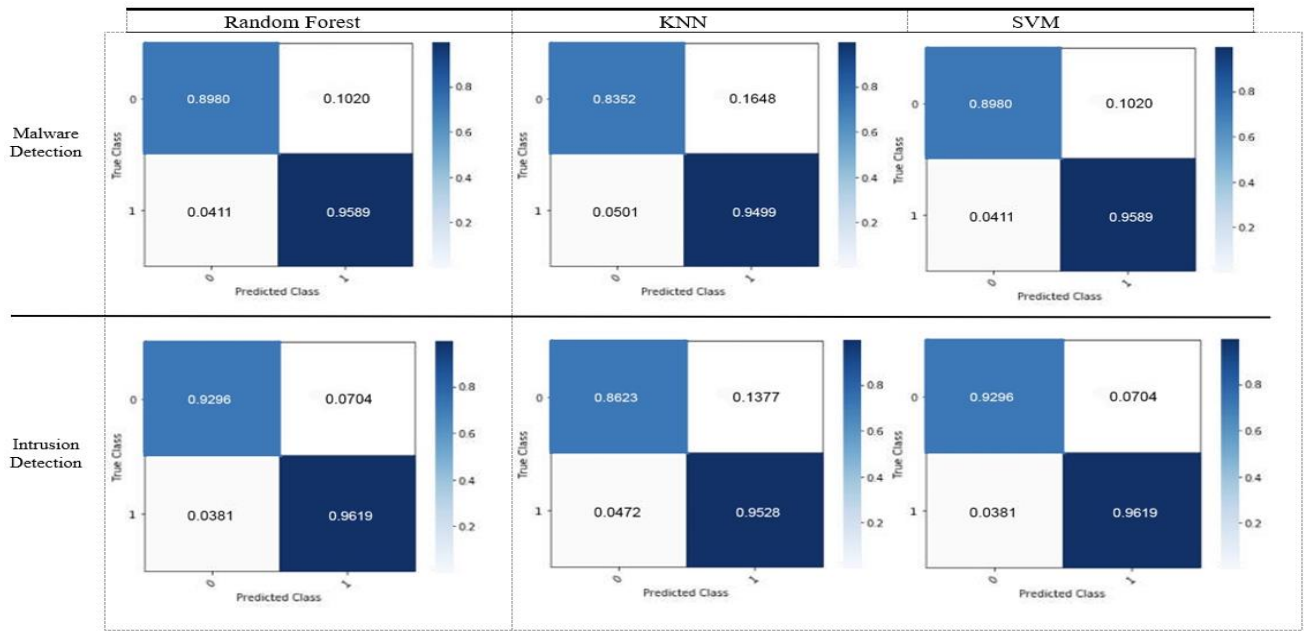

**Figure 3.** Confusion matrix comparison

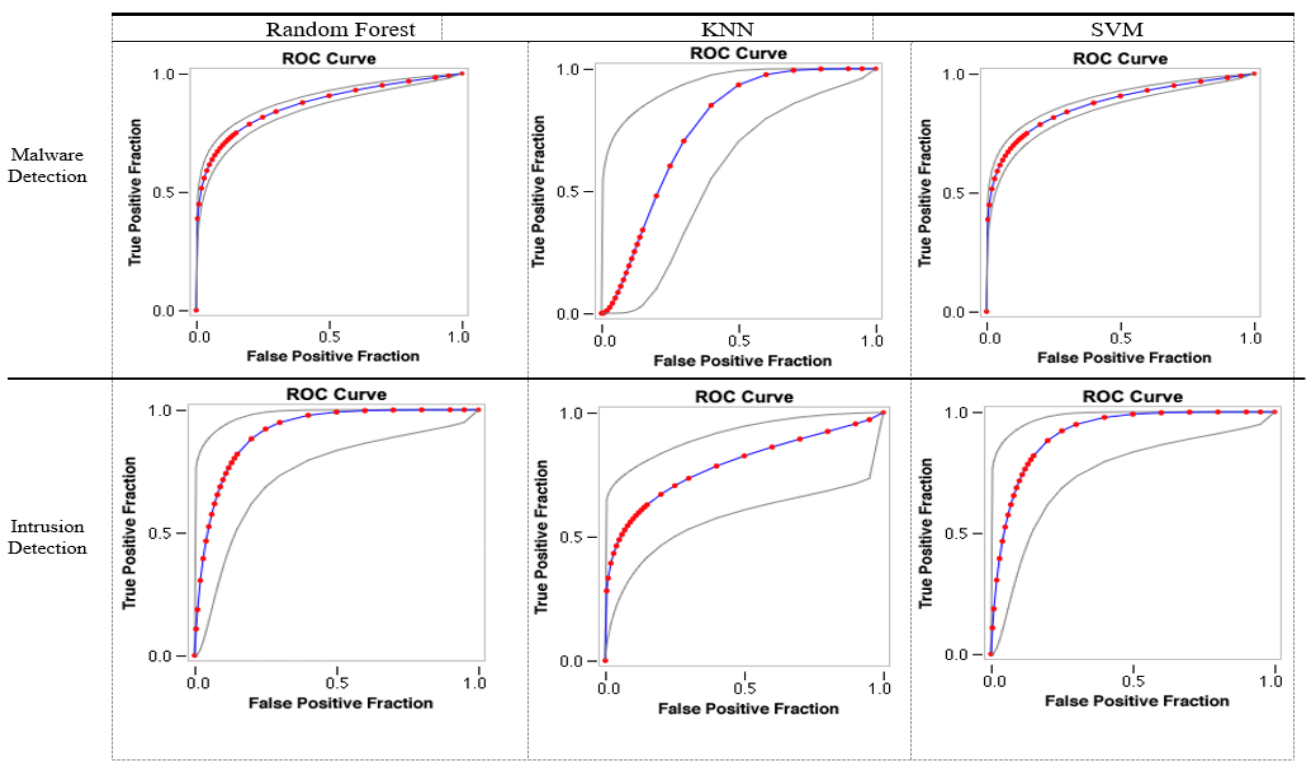

**Figure 4.** ROC curve comparison

# 6. Conclusion

This article has discussed, as noted in Table 5, and tested the effectiveness of three machine learning algorithms used in IoT cybersecurity for malware and intrusion detection trained on a modern dataset. This study highlighted the importance of maintaining and evolving the cybersecurity protocols for IoT devices and networks to prevent malicious attackers. Some limitations in our study include testing more SVM kernels, such as polynomial kernels, to better optimize SVM results. We believe that future kernel testing may lead to much greater accuracies with the SVM for malware detection. This could also address the aforementioned ROC curve inconsistencies. A small improvement for future studies that may be overlooked is combing datasets to allow for more diversity in networking setups, devices, and protocols to allow for more variables to be tested against. Larger datasets may also more accurately represent the entropy present in real-world implementations. Another limitation is the age of the data within the dataset this article used. Even though the dataset was published in 2020, some of the captures within the dataset date back as far as 2018. This presents the issue of trained models lagging up to three years behind current IoT networks. A three-year window allows for many changes to IoT devices like firmware changes, OS version updates, hardware revisions, and new protocols; all of which can lead to potential exploits requiring constant supervision of cybersecurity measures in place to maintain their effectiveness. Our study explored RFs efficacy with modern IoT data and confirmed that it continues to be the most optimal algorithm to use for IoT cybersecurity. While the KNN and SVM did not perform better than the RF, they can still be effectively used for malware and intrusion detection in today's IoT networks.

Table 5. Comparison in Existing Results

| Article | Methodologies | Results | Comparison to Our Results |
|---|---|---|---|
| [1] | Collected data from multiple IoT and non-IoT devices, extracted a set of features from the sessions, then trained, optimized, and tested several classifiers using RF | RF correctly identifies a session's device of origin 99.28% of the time using only the first optimized number of consecutive sessions | RF similarly outperformed all other classifiers in [1] as in our results, however they achieved a higher accuracy compared to our accuracy rate of 92.96% |
| [2] | Collected data from 9 different IoT devices, extracted a set of features from the sessions and then trained, optimized, and tested a classifier using RF with the data. | Identified a session's device of origin with an accuracy of 99% and correctly identified a session as unknown 96% of the time using a window of 20 sessions. | RF was found to be most accurate in both studies. This study can be used to increase RD accuracies in our results |
| [4] | Collect data from three different IoT devices and data from a simulated DoS attack to the dataset. Then, extract features and train and test the five classifiers with the data. | All the classifiers reached accuracy rates of 99.9% except for the Linear SVM with 99.1%. RF performed the best both with and without stateful features. | The highest RF accuracy our results obtained was 98% while the results in [4] were over 99%. Reducing our feature set may help increase accuracy. |
| [6] | Uses a SVM as to not use physical keys. A physical layer authentication scheme using machine learning to improve spoofing detection | Hash functions can be used to speed up identification and game theory could be used to increase accuracy over 90% | Our SVM was found to be the least accurate with 86.23% and 83.52% in intrusion and malware detection, respectively. It boasts speed improvements at the cost of added complexity |
| [13] | Used a simulated dataset to extract and train five classifiers with the data to identify whether its benign or malicious, as well as other attacks | ANN, RF, and DT all reached accuracies of 99.4% and F1-Scores of 0.99. | Our highest accuracy was 92.96% with the RF selecting 6 features. Our sets could be reduced again from 6 to 5 features to approach similar accuracies. |